\documentclass[nolinenumbers]{aastex631}

\usepackage{amsmath}

\submitjournal{ApJ}

\begin{document}

%\title{Generalized equilibrium solution of Fishbone and Moncrief torus for extended GRMHD simulations}

\title{Revisited equilibrium solution of Fishbone and Moncrief torus for extended GRMHD simulations}

\author[0000-0001-8213-646X]{Akhil Uniyal}
\affiliation{Tsung-Dao Lee Institute, Shanghai Jiao Tong University, Shengrong Road 520, Shanghai, 201210, People’s Republic of China}

\author[0000-0002-4064-0446]{Indu K. Dihingia}
\affiliation{Tsung-Dao Lee Institute, Shanghai Jiao Tong University, Shengrong Road 520, Shanghai, 201210, People’s Republic of China}

\author[0000-0002-8131-6730]{Yosuke Mizuno,}
\affiliation{Tsung-Dao Lee Institute, Shanghai Jiao Tong University, Shengrong Road 520, Shanghai, 201210, People’s Republic of China}
\affiliation{School of Physics and Astronomy, Shanghai Jiao Tong University, 800 Dongchuan Road, Shanghai, 200240, People’s Republic of
China}
\affiliation{Institut f\"{u}r Theoretische Physik, Goethe-Universit\"{a}t Frankfurt, Max-von-Laue-Strasse 1, D-60438 Frankfurt am Main, Germany}

\correspondingauthor{Akhil Uniyal, Indu K. Dihingia, Yosuke Mizuno}
\email{akhil\_uniyal@sjtu.edu.cn}
\email{ikd4638@sjtu.edu.cn}
\email{mizuno@sjtu.edu.cn}

\begin{abstract}
Accretion physics has become more important recently due to the detection of the first horizon-scale images of the super-massive black holes of M\,87$^*$ and Sgr~A$^*$ by the Event Horizon Telescope (EHT). General relativistic magnetohydrodynamic (GRMHD) simulations of magnetized accretion flows onto a Kerr black hole have been used to interpret them. However, further testing the theory of gravity by using horizon-scale images requires performing consistent GRMHD simulations in non-Kerr spacetime. In this paper, we revisited the hydrodynamical equilibrium solution of the Fishbone and Moncrief (FM) torus  that can be used to study any stationary, axisymmetric, vacuum, or non-vacuum spacetime. Further, we check the stability of the FM torus in non-Kerr spacetime by general relativistic hydrodynamic simulations. We find that FM torus in non-Kerr spacetime is indeed stable under long-term evolution. We conclude that the generalized FM torus solution would be very useful for creating new GRMHD libraries in extended Kerr black holes.
\end{abstract}


\section{Introduction} \label{sec:intro}

General Relativity (GR) has been widely accepted due to its profound predictions and observational verification \citep[e.g.,][]{Will:2014kxa}. More recently, the detection of the first gravitational waves by the Laser Interferometer Gravitational-Wave Observatory (LIGO) from the binary black hole (BH) merger \citep{LIGOScientific:2016aoc} and the images of the supermassive BH by the Event Horizon Telescope (EHT) at the center of the radio galaxy M\,87 \citep{EventHorizonTelescope:2019dse} and at the center of the Milky Way Sgr~A$^*$ \citep{EventHorizonTelescope:2022wkp} established GR to be one of the well-defined theories of gravity. However, the GR itself has seen some fundamental issues in the observations and theoretical perspectives, such as the existence of the singularity at the center of the BH spacetime \citep{Hawking_Ellis_1973}, the accelerated expansion of the Universe \citep{SupernovaSearchTeam:2001qse}, and the renormalization of the theory of GR \citep{tHooft:1974toh, Goroff:1985th}. Therefore, testing the theory of gravity around compact objects is very important to clarify it.

The general relativistic magnetohydrodynamic (GRMHD) simulation has been an essential tool in the study of the dynamics of astrophysical phenomena near the extreme gravitational field of the compact object in relativistic regimes \citep[e.g.,][]{Banyuls:1997zz, Koide1999RelativisticJF, Gammie:2003rj, DeVilliers:2003gr}. It has been used to study numerous astrophysical phenomena, including the formation of relativistic jets, winds, hot spots, the structure of the hot accretion flow around the BH, and neutron star systems \cite[e.g.,][]{narayan2012grmhd, Liska:2017alm, Parfrey:2017nby, Nathanail:2018dcl, Davis-Tchekhovskoy2020, Porth:2020txf, Dihingia-etal2021, Mizuno-etal2021,Begelman:2021ufo, Dihingia:2022aoc, Chatterjee:2022mxg, Mizuno2022, Das:2023vow, Dihingia-Fendt2024}. The EHT has also provided a large library of 3D GRMHD simulations of the magnetized hot accretion flows in the stationary, axisymmetric spinning vacuum solution of the Einstein field equation: Kerr spacetime \citep{EventHorizonTelescope:2019pgp, EventHorizonTelescope:2022urf}. We can test GR by comparing the synthetic images produced by these libraries with the EHT observed images \citep{EventHorizonTelescope:2019ggy, EventHorizonTelescope:2022xqj}. A diverse work on the semi-analytical model has also been studied to test the GR \citep[e.g.,][]{Narayan:2019imo, EventHorizonTelescope:2020qrl, EventHorizonTelescope:2021dqv, Bauer:2021atk, Ozel:2021ayr, Ayzenberg:2022twz, Younsi:2021dxe, Vagnozzi:2022moj, Uniyal:2022vdu, Uniyal:2023ahv, Jiang:2023img}. However, further testing of the theory of gravity is needed to go beyond Kerr spacetime. The simulation community as well as EHT have already made some efforts in this direction \citep[e.g.,][]{Mizuno:2018lxz, Olivares:2018abq, Fromm:2021flr, EventHorizonTelescope:2022xqj, Nampalliwar:2022ite, Roder:2023oqa}. These simulations and calculations show that images have high degeneracy when it comes to the deformation in Kerr spacetime. Therefore, further studies of horizon-scale dynamics and imaging are inevitable \citep[e.g.,][]{Chatterjee:2023rcz, Chatterjee:2023wti, Jiang:2024vgn}. 

In order to perform the GRMHD simulations in non-Kerr spacetime, we first need to build initial conditions that are consistent with them. It depends on spacetime, and therefore constructing the initial setup, such as the hydrodynamical equilibrium torus in a general framework is an essential requirement. So far, in the GRMHD simulations of magnetized accretion flow around Kerr black holes, most of the work has been done by considering the Fishbone and Moncrief (FM) torus solution \citep{Fishbone1976RelativisticFD} which is applicable not only to Kerr spacetime but also to any stationary and axisymmetric spacetime in GR. In this paper, we apply the idea of the FM torus to any general stationary and axisymmetric spacetime and show the stability of the torus. Previously, initial conditions of equilibrium torus for non-Kerr spacetimes were provided by \cite{Font:2002bi} which is commonly known as FD torus. Although the FD torus has already been used for several GRMHD simulations of non-Kerr spacetime \citep{Mizuno:2018lxz, Gimeno-Soler:2018pjd, Olivares:2018abq, Teodoro:2021ezj, Gimeno-Soler:2021ifv, Roder:2023oqa, Cassing:2023bpt}, the FM torus is widely used to perform GRMHD simulations in Kerr spacetime. It is important to note that other stable tori solutions also exist in the literature such as magnetized tori with a toroidal magnetic field \citep{Komissarov:2006nz}, tori with non-constant angular momentum \citep{DeVilliers:2003gr, Gimeno-Soler:2017qmt, Lei:2008ui} and tori with the alternative initial conditions \citep{Penna:2013zga}. Therefore, building stationary FM tori in generic stationary and axisymmetric backgrounds could provide an alternative framework to study accretion flow in non-Kerr spacetimes. 

To test the generic FM torus solution in non-Kerr spacetime, we perform general relativistic hydrodynamic (GRHD) simulations in the Johannsen-Psaltis (JP) parametrized metric \citep{Johannsen:2011dh, Johannsen:2013szh, Johannsen:2013vgc} which deforms the Kerr metric by the deformation parameters. We show that the general FM torus is stable in long-term evolution, which is a desired characteristic of initial conditions.

This paper is structured in the following ways: We start with the mathematical formulation of the general FM torus in Section~\ref{sec:maths}. We then show the stability of the general FM torus by GRHD simulations in Section~\ref{sec:stability} and conclude our discussion in Section~\ref{sec:conclusion}. We perform the simulations in geometrized units: the speed of light, $c = 1$, the gravitational constant, $G = 1$, and the mass of the central black hole, $M=1$ using the GRMHD code \texttt{BHAC} \citep{Porth-etal2017,Olivares-etal2019}. The length and time are expressed in units of $r_g=GM/c^2$ and $t_g=r_g/c$, respectively.

\section{Mathematical Formulation} \label{sec:maths}
We consider a general form of the stationary, axisymmetric, asymptotically vacuum or non-vacuum flat spacetime in spherical polar coordinates $(t,r,\theta,\phi)$ as
\begin{equation}
    ds^2 = -e^{2\nu} dt^2 + e^{2\psi} (d\phi - \omega dt)^2 + e^{2\mu_1} dr^2 + e^{2\mu_2} d\theta^2,
\end{equation}
where, $\nu$, $\psi$, $\omega$, $\mu_1$, and $\mu_2$ are the function of the coordinates. In this case, all the coefficients can be written in terms of the metric components in the Boyer-Lindquist coordinates as,
\begin{equation}
\begin{aligned}
    e^{2\nu}=\left( \frac{g_{t\phi}}{g_{\phi \phi}}-g_{tt}\right), \quad & \omega = \frac{-g_{t\phi}}{g_{\phi \phi}}, \quad & e^{2\psi}=g_{\phi\phi}, \quad & e^{2\mu_1} = g_{rr}, \quad & e^{2\mu_2} = g_{\theta \theta}.
\end{aligned}
\end{equation}

where all the metric components depend on the radial coordinate $r$ and polar coordinate $\theta$. To construct the hydrostatic equilibrium torus, we would like to have the conserved angular momentum per unit inertial mass ($l=u_\phi u^t$), pressure ($p$), and the rest-mass density of the matter ($\rho$). We start with the integral form of the relativistic Euler equation for isentropic fluid $dS=0$ \citep{Fishbone1976RelativisticFD},
\begin{equation}
    dW=d[\ln(h)+\nu]=(u_{(\phi)})^2d(\psi-\nu)-u_{(\phi)}[1+(u_{(\phi)})^2]^{1/2}e^{\psi-\nu}d\omega,
\end{equation}
where $u^\rho=(u^t,0,0,u^\phi)$ is the four-velocity of the fluid and $h=(e+p)/\rho$ is the specific enthalpy with total energy density, $e$. In the Newtonian limit, $W$ is the effective potential with contributions from the centrifugal and gravitational forces. 
Therefore, in the hydrodynamic equilibrium, $W$ can be defined as the equation of the balance by the pressure gradient in the following way,
\begin{equation}
    W-W_{\rm in}=-\int_0^P \frac{dp}{w},
\end{equation}
here $w=e+p$. The subscript $"in"$ hereon will represent the value of the physical quantity in the inner edge ($r_{in}$) of the torus at the equatorial plane ($\theta=\pi/2$) where the pressure vanishes.
Using the polytropic equation of state $p=K \rho^\gamma$, where $\gamma$ and $K$, are the adiabatic index and polytropic constant respectively, we can integrate the above equation which gives,
\begin{equation}
    \ln(h)- \ln(h_{\rm in}) =\ln \left( 1 + \frac{\gamma}{\gamma -1} K \rho^{\gamma -1} \right).
\end{equation}
The distribution of density and pressure of the torus are expressed as,
\begin{equation}
\begin{aligned}
    \rho=&\left(\frac{(\gamma-1) (e^{\ln(H)}-1)}{\gamma K}\right)^{\frac{1}{\gamma-1}},\\
    p=&K \left(\frac{(\gamma-1) (e^{\ln(H)}-1)}{\gamma K}\right)^{\frac{\gamma}{\gamma-1}},
\label{eq:dp}
\end{aligned}
\end{equation}
where
\begin{equation}
\ln(H)=\ln(h) -\ln(h_{\rm in}).
\end{equation}
and
\begin{equation}\label{eq:8}
\ln(h)=-\frac{1}{2} \sqrt{\frac{4 l^2 e^{2\nu}}{e^{2\psi}}+1}+\frac{1}{2} \ln \left(\frac{\sqrt{\frac{4 l^2 e^{2\nu}}{e^{2\psi}}+1}+1}{e^{2\nu}}\right)-l \omega + W_{in},
\end{equation}
Note that the equation of state here considered is the polytropic equation of state which satisfied $\displaystyle{\lim_{P \to 0}h= \displaystyle{\lim_{P \to 0} \frac{\rho + P}{n}} = 1}$, $n$ is the baryon number density. Hence, the boundary of the disk will be the surface where $\ln(h)=0$ and therefore here $\ln(h_{in})=0$ \citep{Fishbone1976RelativisticFD}. This further helps in calculating $W_{in}$ which will be the first three terms in equation (\ref{eq:8}) at $r=r_{in}$ and $\theta=\pi/2$. The angular momentum per unit inertial mass $l$ can then be expressed as the solution of the pressure gradient equation ($[\ln(h_{\rm })]_{,r}=0$ at $\theta=\pi/2$) following \citep{Fishbone1976RelativisticFD},
\begin{equation}
[\ln(h_{\rm })]_{,r}=-\frac{1}{2}(\psi+\nu)_{,r}-l\omega_{,r}+\frac{1}{2}(\psi-\nu)_{,r}\left(1 + \frac{4l^2e^{2\nu}}{e^{2\psi}} \right)^{1/2},
\end{equation}
Here $,r$ denotes the partial derivative with respect to radial coordinate ($r$) at $\theta=\pi/2$. The equation gives two solutions for the angular momentum per unit inertial mass, one representing the counter-rotating disk and the other for the co-rotating disk. The co-rotating disk solution leads to the following expression:
\begin{equation}
l=A/B,
\label{eq:ang}
\end{equation}
where
\begin{equation}
\begin{aligned}
        A= & -g_{t\phi} \left\{g_{\phi \phi } \left[g_{tt,r} g_{\phi \phi , r}+2 \left(g_{t\phi,r} \right)^2\right]+g_{tt} \left(g_{\phi \phi ,r} \right)^2\right\}+ \\ 
        &\sqrt{\left[\left(g_{t\phi,r} \right)^2-g_{tt,r} g_{\phi \phi ,r} \right] \left[g_{\phi \phi } \left(g_{\phi \phi } g_{tt,r} - g_{tt} g_{\phi \phi,r } \right)+2 g_{t\phi}^2 g_{\phi \phi,r } -2 g_{t\phi} g_{\phi \phi } g_{t\phi,r} \right]^2}\\ 
        &+g_{\phi \phi }^2 g_{tt},_r g_{t\phi,r} + g_{tt} g_{\phi \phi } g_{t\phi,r} g_{\phi \phi,r } + 2 g_{t\phi}^2 g_{t\phi,r} g_{\phi \phi,r } ,\\
        B= & g_{tt,r} \left(g_{\phi \phi }^2 g_{tt,r}+4 g_{t\phi}^2 g_{\phi \phi,r }-4 g_{t\phi} g_{\phi \phi } g_{t\phi,r} \right)- \\
        &2 g_{tt} \left\{g_{\phi \phi } \left[g_{tt,r} g_{\phi \phi,r } -2 \left(g_{t\phi,r} \right)^2\right]+2 g_{t\phi} g_{t\phi,r} g_{\phi \phi,r } \right\}+g_{tt}^2 \left(g_{\phi \phi,r } \right)^2.
\end{aligned}
\end{equation}

The above equation is the generalized version of Eq.$(3.8)$ of Fishbone and Moncrief for a generic stationary and axisymmetric spacetime \citep{Fishbone1976RelativisticFD}. The four-velocity components are also needed to construct the complete structure of the torus. First, we set the radius of the maximum density of the torus ($r_{\rm max}$). By supplying $r_{\rm max}$ to Eq.~(\ref{eq:ang}), we obtain the constant angular momentum for the torus. It is important to note that since the pressure is required to increase outwards from $r_{\rm in}$, i.e., $[\ln(h_{\rm})]_{,r}>0$ at $r=r_{\rm in}$, we can not fix the angular momentum at $r=r_{in}$. This further restricts the inner edge of the disk to be $r_{\rm in}>r_{\rm ph_+}$, where $r_{\rm ph_+}$ is the retrograde photon radius of the circular photon orbit. Note that for $r<r_{\rm ph_+}$, the formation of the circular orbits needs acceleration due to the pressure gradient outwards. However, at $r=r_{\rm in}$, it is directed inwards. Such criteria also prevent the disk from having any cusp at the inner edge \citep{Kozlowski1978, Font:2002bi}. In the case of the massive particle, as it approaches the photon sphere, it moves more and more rapidly, which results in the divergence of angular momentum and energy per unit mass. Therefore, the positive root of $B=0$ will provide the photon sphere radius. A general analytical expression is difficult to find, thus we provide the prograde and retrograde photon sphere radius for the lowest order parameter variation of the JP metric in Tabel~\ref{tab:ph}.

\begin{deluxetable}{cccc}
\tablecaption{Photon Sphere radius for the JP metric with black hole spin 0.9375 and $\theta=\pi/2$. The lowest order parameters with one varying at a time.\label{tab:ph}}
\tablewidth{0pt}
\tablehead{
\colhead{JP parameter} & \colhead{Prograde Photon Sphere Radius $r_{ph_-}$ (M)} & \colhead{Retrograde Photon Sphere Radius $r_{ph_+}$ (M)}
}
\startdata
$Kerr$(no deformation)  & 1.43452 & 3.94412 \\
$\alpha_{13}=0.5$  & 1.55483 & 3.99472 \\
$\alpha_{22}=0.5$  & 1.37201 & 4.04279 \\
$\alpha_{52}=0.5$  & 1.43452 & 3.94412 \\
$\epsilon=0.5$  & 1.43452 & 3.94412 \\
\enddata
\end{deluxetable}

Further, we use this angular momentum to calculate the initial conditions of the four-velocity components. We can calculate the azimuthal component of the contravariant four-velocity by the following transformations:
\begin{equation}
    u^{\phi} = \left( \frac{e^{\nu} u_{(\phi)}}{e^{\psi}\sqrt{1+u_{(\phi)}^2}} + \omega \right) u^t,
\label{eq9}
\end{equation}
where $u_{(\phi)}$ is the azimuthal velocity component in the locally non-rotating frame (LNRF) and the basis vectors in this frame can be transformed as \citep{Bardeen1972RotatingBH},
\begin{equation}
    \begin{aligned}
        e_{(t)}=&e^{-\nu}\left( \frac{\partial}{\partial t} + \omega \frac{\partial}{\partial \phi} \right), ~~
        e_{(r)}=e^{-\mu_1}\frac{\partial}{\partial r},~~
        e_{(\theta)}=e^{\mu_2}\frac{\partial}{\partial \theta},~~
        e_{(\phi)}=e^{-\psi}\frac{\partial}{\partial \phi}.
    \end{aligned}
\end{equation}
Therefore, with the help of the above transformations, we can write the four-velocity transformations as follows \citep{Bardeen1972RotatingBH},
\begin{equation}
    \begin{aligned}
        u_{(\phi)}=&e^{-\psi}u_{\phi},\
        u_{(t)}=e^{-\nu}(u_t+\omega u_\phi)=-\sqrt{1+(u_{(\phi)})^2},\
        u_{(t)}=-e^{\nu}u^t,
    \end{aligned}
\end{equation}
which reduce $u^{\phi}$ to be,
\begin{equation}
    u^{\phi}=\frac{u_{(\phi)}}{e^{\psi}}+\frac{\omega}{e^\nu}\sqrt{1+(u_{(\phi)})^2}.
\end{equation}
Following \cite{Fishbone1976RelativisticFD}, $u_{(\phi)}$ can be expressed as,
\begin{equation}
    u_{(\phi)}=\sqrt{\frac{-1+\sqrt{1+4l^2e^{2(\nu-\psi)}}}{2}}.
\end{equation}
The time-component of the contravariant four-velocity, $u^t$ can be calculated using the normalization condition $g_{\alpha \beta}u^{\alpha}u^{\beta}=-1$ where $u^r=u^{\theta}=0$ while constructing the torus. Therefore, the azimuthal component of the four-velocity of the torus can be constructed using Eq.~(\ref{eq9}) and the above-constructed relations. The pressure can be calculated within the torus by using Eq.~(\ref{eq:dp}). Therefore, the structure of the torus has been formulated in this section. We will test the stability of the constructed generic FM torus in the next section (Sec.~\ref{sec:stability}).

\section{General Relativistic Hydrodynamic Simulation Tests}\label{sec:stability}

In the previous section, we described the procedure to construct general FM torus in non-Kerr spacetime.
In this section, we test the stability of the newly constructed general FM torus through numerical simulations. 
To represent the non-Kerr metric, we use the Johannsen-Psaltis (JP) parametrized metric in the Boyer-Lindquist (BL) coordinates, which is given as \citep{Johannsen:2013szh}:
\begin{equation}
\begin{aligned}
    ds^2 = & g_{tt}dt^2 + 2g_{t\phi}dtd\phi + g_{rr}dr^2 + g_{\theta\theta}d\theta^2 + g_{\phi\phi}d\phi^2.
\end{aligned}
\end{equation}
the non-zero metric components are given by,
\begin{equation}
\begin{aligned}
    g_{tt} = & -\frac{\Tilde{\Sigma}\left(\Delta - A_2^2a^2\sin^2{\theta} \right)}{F},~~~
    g_{t\phi} = -\frac{a\Tilde{\Sigma}\left( (r^2+a^2)A_1A_2-\Delta\right)\sin^2\theta}{F},\\
    g_{rr} = & \frac{\Tilde{\Sigma}}{\Delta A_5}, ~~~
    g_{\theta\theta} = \Tilde{\Sigma},~~ {\rm and} ~~
    g_{\phi\phi} = \frac{\Tilde{\Sigma}\left((r^2+a^2)^2A_1^2 - a^2\Delta\sin^2\theta\right)\sin^2\theta}{F},
\end{aligned}
\end{equation}
where $A_1$, $A_2$, and $A_5$ are the function of radial coordinate ($r$) only and,
\begin{equation}
\begin{aligned}
    \Tilde{\Sigma}= & r^2 + a^2 \cos^2\theta, ~~~
    \Delta = r^2 - 2Mr + a^2 + \sum_{n=3}^{\infty} \epsilon_n \frac{M^n}{r^{n-2}},\\
    F = &\left((r^2 + a^2)A_1 - a^2A_2\sin^2\theta\right)^2, ~~~
    A_1 = 1 + \sum_{n=3}^{\infty}\alpha_{1n} \left(\frac{M}{r}\right)^n,\\
    A_2 = & 1 + \sum_{n=2}^{\infty}\alpha_{2n} \left(\frac{M}{r}\right)^n, ~~{\rm and}~~
    A_5 = 1 + \sum_{n=2}^{\infty}\alpha_{5n} \left(\frac{M}{r}\right)^n.
\end{aligned}
\end{equation}
The coefficients $\epsilon_n, \alpha_{1n}, \alpha_{2n}$, and $\alpha_{5n}$ are the deviation parameters. For our purposes, we stick to the first-order contribution of the deformation parameters and vary one of these parameters in simulations with $\epsilon_3 \equiv \epsilon$. It is important to note that, as the deformation parameter vanishes, it returns to the Kerr metric.

We perform 2D GRHD simulations of general FM torus in Kerr and non-Kerr spacetimes on a spherical polar grid where the grid spacing is logarithmic in the radial direction and linear in the polar direction. We implemented the horizon penetrating form of the JP metric in the \texttt{code BHAC} \citep{Porth-etal2017} using Eq.~(111) of \citep{Johannsen:2013szh} to represent the non-Kerr spacetime. 

As an initial condition, we apply the general FM torus, which is in hydrostatic equilibrium. We use the Kerr metric and the metric with deformation from Kerr spacetime. In order to make a fair comparison, we fix the total mass content within the torus. Here we choose the inner radius of the torus at $r_{in}=6\,r_g$ and the density maximum radius at $r_{\rm max}=12\,r_g$ which provides the angular momentum. In that case, the mass content inside the torus remains the same. To check the stability of the torus, we add a random perturbation in the pressure up to $4 \%$. In all the simulation runs, we use the same random seed number so that we can have the same initial setup. The simulation here is performed for black hole spin $a=0.9375$, adiabatic index $\gamma=5/3$, and polytropic constant $K=0.001$.

We treat the outside of the torus as an atmosphere that is filled with very low-density matter. We fix the velocity in the Eulerian frame to be zero. Pressure and density set the very low values of $\rho=\rho_{\rm min} r^{-1.5}$ and $p=p_{\rm min} r^{-2.5}$, where $\rho_{\rm min}=10^{-5}$ and $p_{\rm min}=10^{-7}$ respectively, to avoid the unphysically small or negative values during the simulation run. 

The effective grid resolution of the simulations is $1024 \times 512$ grids with two levels of static mesh refinement. At the inner and outer radial boundaries, we apply standard inflow/outflow boundary conditions by copying the physical quantities. At the polar boundaries, reflecting boundary conditions are applied. For azimuthal direction, periodic boundary conditions are employed.
 
\begin{figure*}[t]
    \centering
  \includegraphics[width=0.9\textwidth]{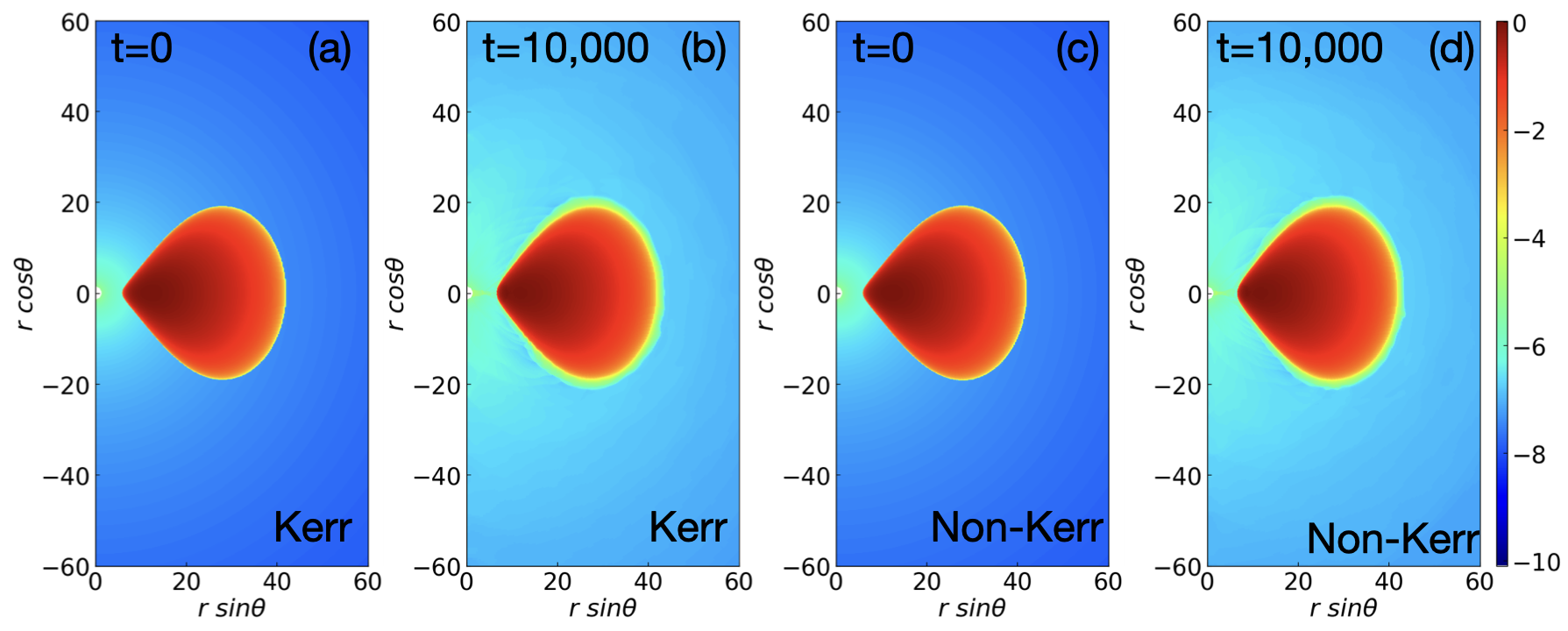}
    \caption{Logarithmic density distribution with at $t=0\,t_g$ (a) and $t=10,000\,t_g$ (b) for Kerr metric case and $t=0\,t_g$ (c) and $t=10,000\,t_g$ (d) for non-Kerr metric case with $\alpha_{22}=0.5$.}
    \label{comp}
\end{figure*}

Fig.~\ref{comp} shows the distribution of the density for the Kerr spacetime (panels \ref{comp}(a) and \ref{comp}(b)) and non-Kerr spacetime with the deformation $\alpha_{22}=0.5$ (panels \ref{comp}(c) and \ref{comp}(d)) at $t=0$ and $t=10,000\,t_g$. We note that for the non-Kerr spacetime case, we set zero for the other deformation parameters. From the $2D$ density distributions, it is clearly shown that the torus remains stable, but the matter outside of the torus falls onto a BH for long-term evolution. We expect that some small changes may happen due to the given initial perturbation within the torus. Nevertheless, the results remain similar in both cases. We found similar results for all non-Kerr spacetime cases by adding different deformation parameters. We do not display them explicitly here to avoid repetition. 

\begin{figure*}[h!]
    \centering
    \includegraphics[width=\textwidth]{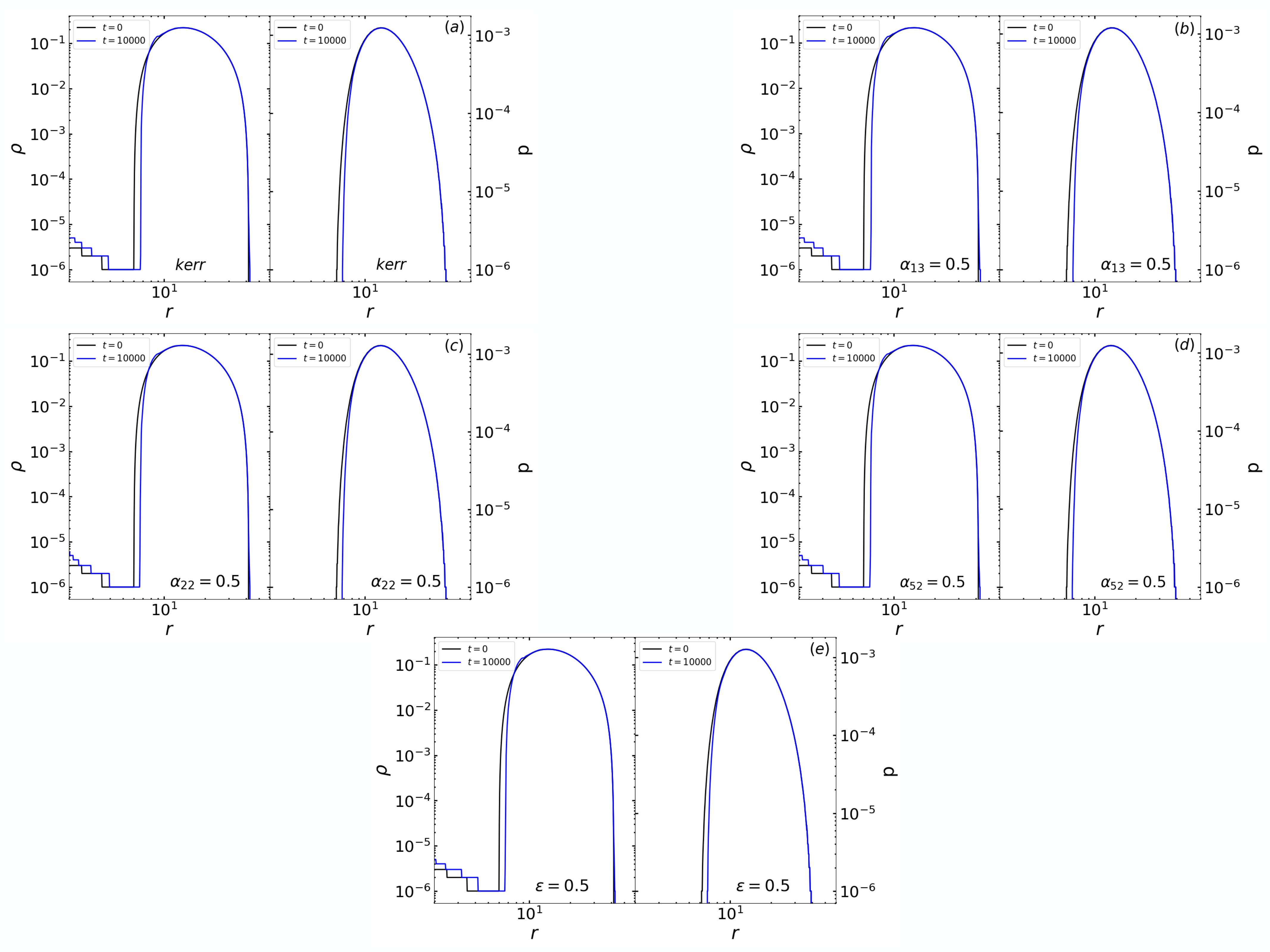}
    \caption{1D profiles of vertically-integrated radial density ({\it left}) and pressure ({\it right}) at $t=0$ (black) and $t=1,0000$ $t_g$ (blue) for Kerr spacetime ((a)), non-Kerr spacetime with $\alpha_{13}=0.5$ ((b)), with $\alpha_{22}=0.5$ ((c)), with $\alpha_{52}=0.5$ ((d)), and with $\epsilon=0.5$ ((e)).}
    \label{fig:radial}
\end{figure*}

We further check the stability of the torus structure quantitatively by the vertically integrated radial density and pressure profiles, which are shown in Fig.~\ref{fig:radial}. We varied the one deformation parameter ($a_{13}$, $a_{22}$, $a_{52}$, or $\epsilon$) for each non-Kerr case and compared their radial density and pressure profiles at $t=0\,t_g$ (solid black line) and $t=10,000\,t_{g}$ (solid blue line). The values of the non-zero Kerr parameters are written on each panel. We observed that the main torus profiles remain stable in the long-term evolution, i.e., unchanged in their initial structure even at time $t=10,000\,t_g$. However, due to the small perturbation in the torus, we see some small changes at the surface of the torus in all the cases, including the case for Kerr spacetime (Fig.~\ref{fig:radial}a).

\begin{figure}[h]
    \centering
    \includegraphics[scale=0.35]{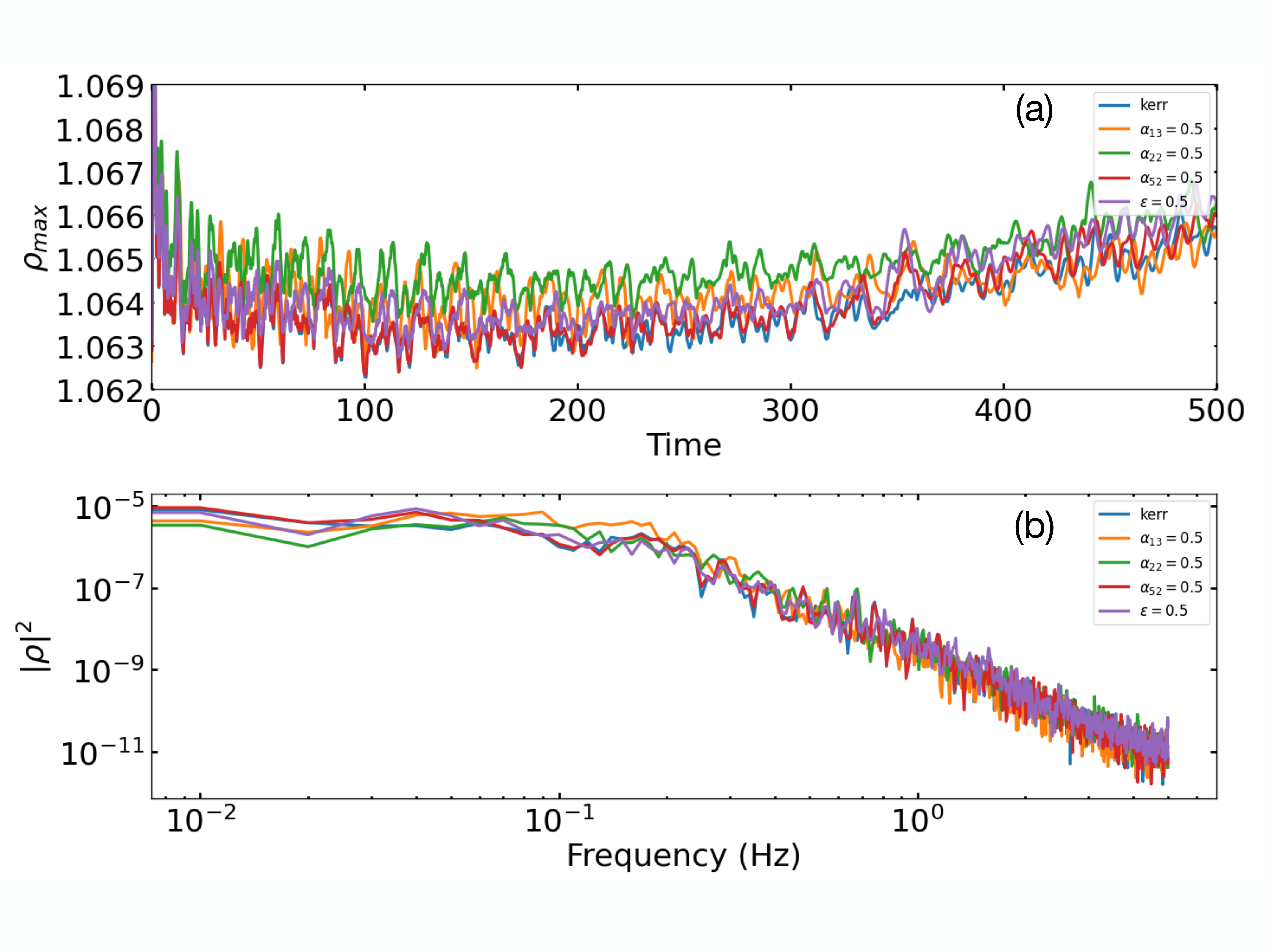}
    \caption{Maximum density with time in units of $t_g$ (panel (a)) and corresponding power spectral density (panel (b)) for Kerr, $\alpha_{13}=0.5$, $\alpha_{22}=0.5$, $\alpha_{52}=0.5$, and $\epsilon=0.5$.}
    \label{fig: psd}
\end{figure}

The final confirmation of the stability check comes from the oscillation of maximum density $(\rho_{\rm max})$ within the torus caused by the initial perturbation, as shown in Fig.~\ref{fig: psd}. In panel (a), we present the evolution of $\rho_{\rm max}$ with time, where we observe the random small oscillations clearly that decay (damping) in long-term evolution for all cases. Even up to $t=10,000\,t_g$, we see such random oscillations. The value of $\rho_{\rm max}$ increases with time since all the perturbed high-pressure regions started to move to the location of the density maximum to make the torus equilibrium. The general trend has not changed in the different spacetimes. To further characterize these oscillations, we show the corresponding power spectral density (PSD) for different spacetime cases in panel (b). From this figure, we do not observe any dominant peaks in the PSD due to the provided random perturbation. Dominate peaks in PSD usually correspond to quasi-periodic oscillations (QPOs) in the emission profile \citep{Zanotti:2004kp, Avellar2018KilohertzQI}, However, such studies need to be done in the MHD, where the turbulent magnetic field can give rise to more oscillation in the evolution of the initially constructed torus. These results will be reported soon in the upcoming work. The important point is that for all the cases, the PSD is mostly consistent, showing the robustness of the initial torus. Overall, the long-term GRHD simulations show that generic FM torus remains stable and can be used for further studies of magnetized accretion flows in the GRMHD simulations. To make our claim more robust, we study the stability of the generalized FM torus by changing values of $r_{\rm in}$ and $r_{\rm max}$ and also in one more considered metric (Kerr-Sen metric). The results of these studies are shown in Appendix~\ref{AppendixA} and \ref{AppendixB}, respectively.

For the sake of completeness, we compare the initial structures ($t=0$) of two torus solutions, the FD torus and the FM torus for non-Kerr spacetime in Appendix~\ref{AppendixC}. Both torus have the same inner and outer edge radius. However, in the interior of the torus, there are some small deviations. 
We will further expand the study to investigate the influence of these differences by using GRMHD simulations, which will be reported soon.

\section{Summary and Conclusion}\label{sec:conclusion}

Testing the theory of gravity has become more important recently. To understand the dynamics of matter near the central compact objects, GRMHD simulation is one of the key tools. High-resolution $3D$ GRMHD simulations in Kerr spacetime have been used to study the magnetized accretion flows near the supermassive BH in Sgr~A$^*$ and M\,87$^*$ \citep{EventHorizonTelescope:2019pgp, EventHorizonTelescope:2022urf}. However, the testing theory of gravity will need to perform GRMHD simulations in non-Kerr spacetimes. To do it, the initial conditions need to be formed carefully, which is consistent with the spacetime characteristics. In Kerr spacetime, we have widely used one of the hydrostatic equilibrium torus solutions, the so-called FM torus to perform GRMHD simulations. In this paper, we present more general accretion tori solutions using the FM approach in stationary, axisymmetric, non-Kerr spacetimes.

To investigate the stability of our torus solution, we performed 2D GRHD simulations of FM tori in non-Kerr spacetimes. We fixed the inner radius as well as the mass content within the torus for all cases and introduced a $4\%$ of random perturbation to the gas pressure in the torus. The JP parametrized metric is used as an example for the deformation from Kerr spacetime. By checking the density and pressure distribution, we show that the torus remains stable for long-term evolution, even when the deformation parameters are non-zero. Furthermore, we looked into the oscillation of the maximum density within the torus and its corresponding PSD to study the effect of the perturbation on the torus structure. We found that the FM torus in non-Kerr spacetime behaves similarly to the Kerr spacetime. Therefore, we conclude that generic FM tori are stable and ready to apply for the GRMHD simulations of magnetized accretion flows in non-Kerr spacetime. It is important to note that the simulation performed in the paper is only for Kerr-like spacetime however the generality of the procedure is valid in all axisymmetric, stationary, and vacuum spacetimes. Therefore we can apply the same FM tori solution in any non-Kerr spacetime.

In the past, the simulation library of magnetized accretion flows onto a black hole in Kerr spacetime has been well-developed and used to understand the near-horizon physics from the black hole shadow images observed by EHT \citep{EventHorizonTelescope:2019pgp, EventHorizonTelescope:2022urf}. However, the simulation library of magnetized accretion flows onto a black hole beyond GR spacetime has not been much developed yet. Although it is essential for testing the theory of gravity via black hole shadow images. Therefore, the development of a new simulation library using non-Kerr spacetime is an urgent task. It is important to note that a horizon-penetrating form of the most general axisymmetric, stationary, asymptotically flat spacetime to use generic FM torus is needed for the long-term evolution of GRMHD simulations. Recently, several new horizon-penetrating forms of the most general axisymmetric, stationary, asymptotically flat spacetime have been proposed \citep[e.g.,][]{Johannsen:2013szh, Konoplya2021, Kocherlakota:2023vff, Ma:2024kbu}. Here, we computed generic initial conditions for the FM torus solution for any non-Kerr stationary, axisymmetric, asymptotically vacuum, or non-vacuum flat spacetime. Further application work using the MHD simulations will be reported soon.

\begin{acknowledgments}
This research is supported by the National Key R\&D Program of China (No.\,2023YFE0101200), the National Natural Science Foundation of China (Grant No.\,12273022), and the Shanghai Municipality orientation program of Basic Research for International Scientists (Grant No.\,22JC1410600). The simulations were performed on TDLI-Astro and Siyuan Mark-I at Shanghai Jiao Tong University. We appreciate the thoughtful comments and suggestions provided by the anonymous reviewers that have improved the manuscript.
\end{acknowledgments}

\appendix
\section{Stability check for different \lowercase{$r_{in}$} and \lowercase{$r_{max}$}}\label{AppendixA}

The gravity has a strong effect near the horizon; therefore, we check the stability of the torus by considering the inner edge to be $r_{\rm in}=4.5\,r_g$ and the density maximum $r_{\rm max} = 10\,r_g$. The choice of $r_{in}$ is considered such that it remains outside the photon sphere radius $(r_{ph_+})$ as shown in Table~\ref{tab:ph}. We fixed the initial condition that the torus has the same mass content. We use two different JP parameter $\alpha_{13}=0.5$ and $\alpha_{22}=0.5$ with black hole spin $a=0.9375$.

\begin{figure*}[h]
    \centering
  \includegraphics[width=0.9\textwidth]{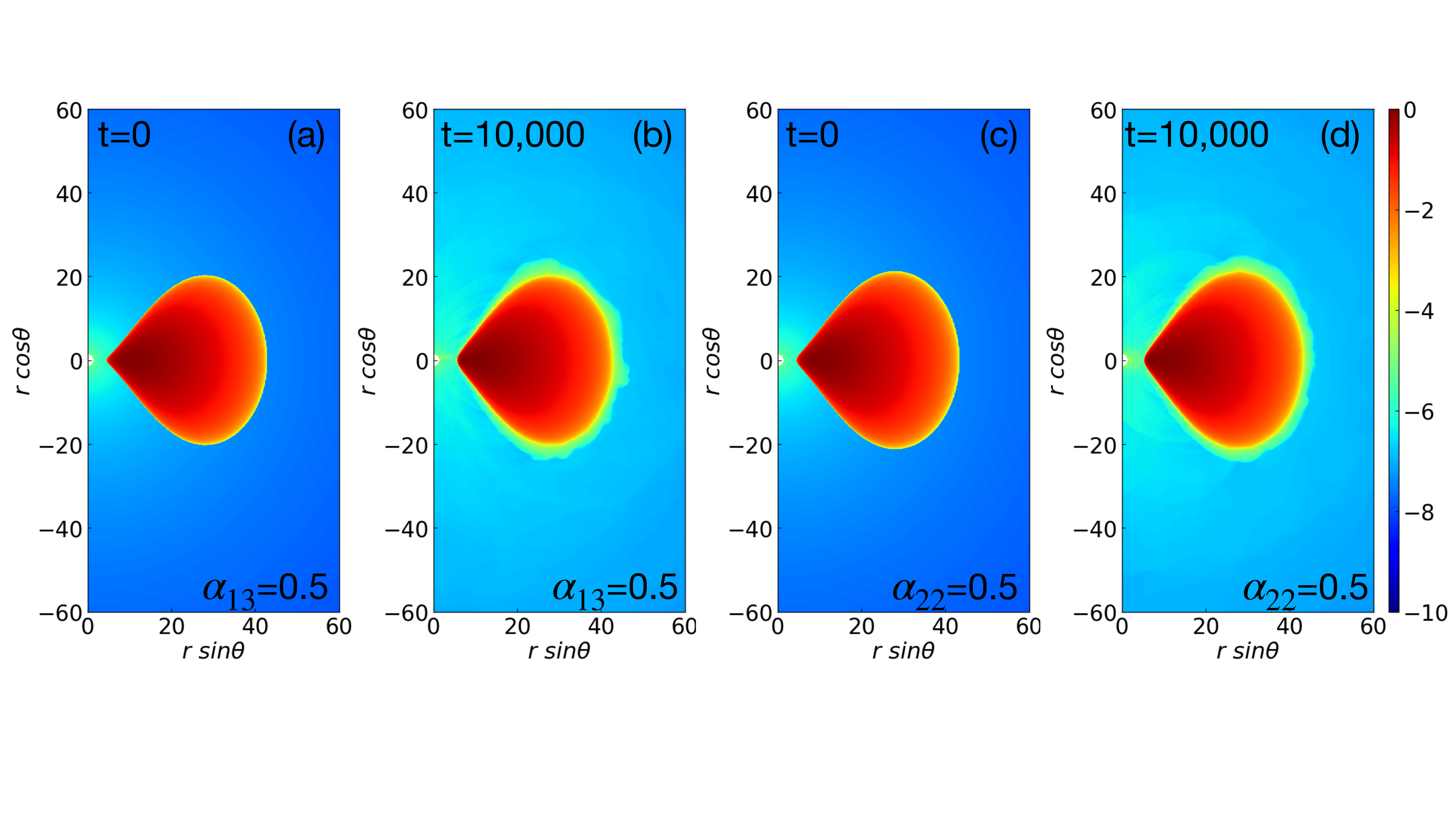}
    \caption{Logarithmic density distribution of the torus with $r_{\rm in}=4.5\,r_g$ and $r_{\rm max} = 10\,r_g$ at $t=0\,t_g$ (a) and $t=10,000\,t_g$ (b) for $\alpha_{13}=0.5$ and $t=0\,t_g$ (c) and $t=10,000\,t_g$ (d) for $\alpha_{22}=0.5$. We used the same mass content within the torus and fixed the black hole spin $a=0.9375$.}
    \label{Fig:lessrin}
\end{figure*}
After time evolution up to $t=10000\,t_g$, the initial and final torus structures are shown in Fig. \ref{Fig:lessrin}. The final structure of the torus indicates that it remains in hydrodynamic equilibrium. However, a loss of a small amount of very low density ($\lesssim 10^{-6}$) matter is observed due to the initial perturbation in the pressure. The main torus structure is unchanged in the long-term evolution. Therefore, we conclude that the generic FM torus can remain hydrodynamically stable even in extreme gravitational environments.

\section{Stability check for the Kerr-Sen metric}\label{AppendixB}

The Kerr-Sen metric in the Boyer-Lindquist (BL) coordinates can be written as \citep{Sen:1992ua},
\begin{equation}
\begin{aligned}
ds^2=-\left(1-\frac{2M r}{\Sigma}\right)dt^2+\Sigma \left(\frac{dr^2}{\Delta}+d\theta^2\right)+\left(\Sigma+a^2\sin^2\theta+\frac{2M r a^2 \sin^2\theta}{\Sigma}\right)\sin^2\theta d\phi^2-\frac{4M r a}{\Sigma}\sin^2\theta dt d\phi\,
\end{aligned}
\end{equation}
where $\Delta=r(r+2b)-2 M r+a^2$ and $\Sigma=r(r+2b)+a^2\cos^2\theta$ with $b=\frac{Q^2}{2}$ depends on the charge $Q$. The horizon of the Kerr-Sen metric is given by $r_\pm=(1-b)\pm\sqrt{\left(1-b\right)^2-a^2}$. We set the same initial torus condition as earlier cases with $r_{\rm in}=6\,r_g$ and $r_{\rm max} = 12\,r_g$. We fixed the spin parameter $a=0.9375$ and simulated for the two different charge cases $Q=0.05$ and $Q=0.1$. We used the horizon penetrating form of JP metric as described in Section~\ref{sec:stability} with $A_2=A_5=1$ and $A_1=\frac{r(r+2b)+a^2}{r^2+a^2}$.

\begin{figure*}[h]
    \centering
  \includegraphics[width=0.9\textwidth]{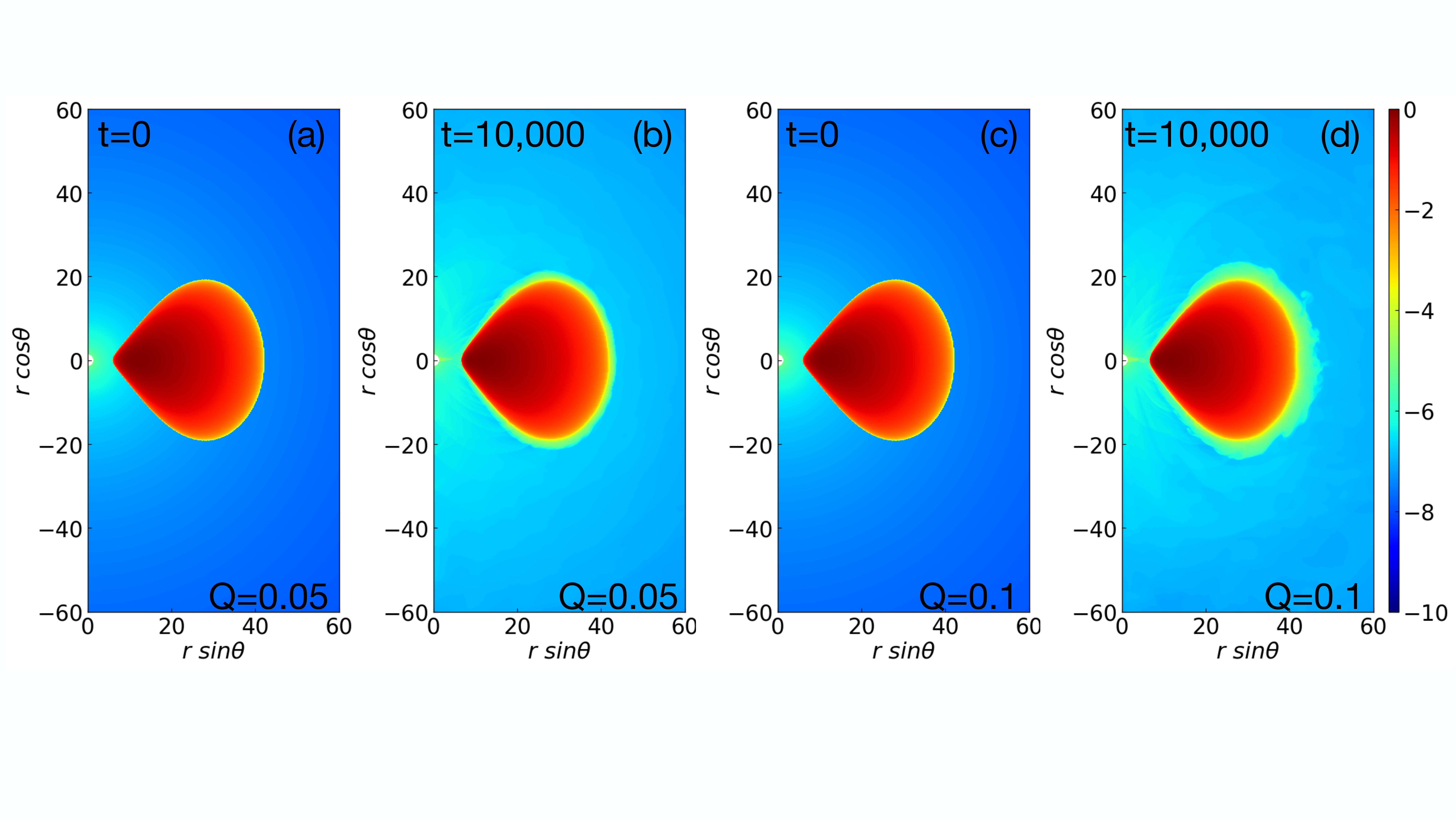}
    \caption{Logarithmic density distribution for the Kerr-Sen metric with at $t=0\,t_g$ (a) and $t=10,000\,t_g$ (b) for charge $Q=0.05$ and $t=0\,t_g$ (c) and $t=10,000\,t_g$ (d) for charge $Q=0.1$. We fixed the black hole spin $a=0.9375$.}
    \label{Fig:Kerr-Sen}
\end{figure*}
The simulation result is shown in Fig.~\ref{Fig:Kerr-Sen}. It can be seen that for a long-term simulation run, the torus remains stable however there is a small amount of matter loose from the torus surface due to the applied perturbation in the pressure. However, the major torus structure is unchanged in the long-term evolution.
Therefore, it can be concluded that the torus solution can be further used to test the other scenario including the magnetic field and other non-Kerr metrics. The further results will be reported in the upcoming work.

\section{Comparison of initial structure between FM and FD torus}\label{AppendixC}

In this section, we compare the initial structure between the two hydrostatic equilibrium solutions, FM and FD torus at $t=0$ for parameterized JP metric with $\alpha_{22}=0.5$ as a representative of non-Kerr spacetime. For FM torus, we use $r_{\rm in}=6\,r_g$ and $r_{\rm max} = 12\,r_g$. To make a fair comparison, we keep the same integrated mass content in the FD torus as that in the FM torus by fixing the angular momentum appropriately ($l_*=-u_\phi/u_t=3.6515$). We also fix the inner radius for FD torus at $r_{\rm in}=6\,r_g$.

\begin{figure}[h]
    \centering
    \includegraphics[scale=0.35]{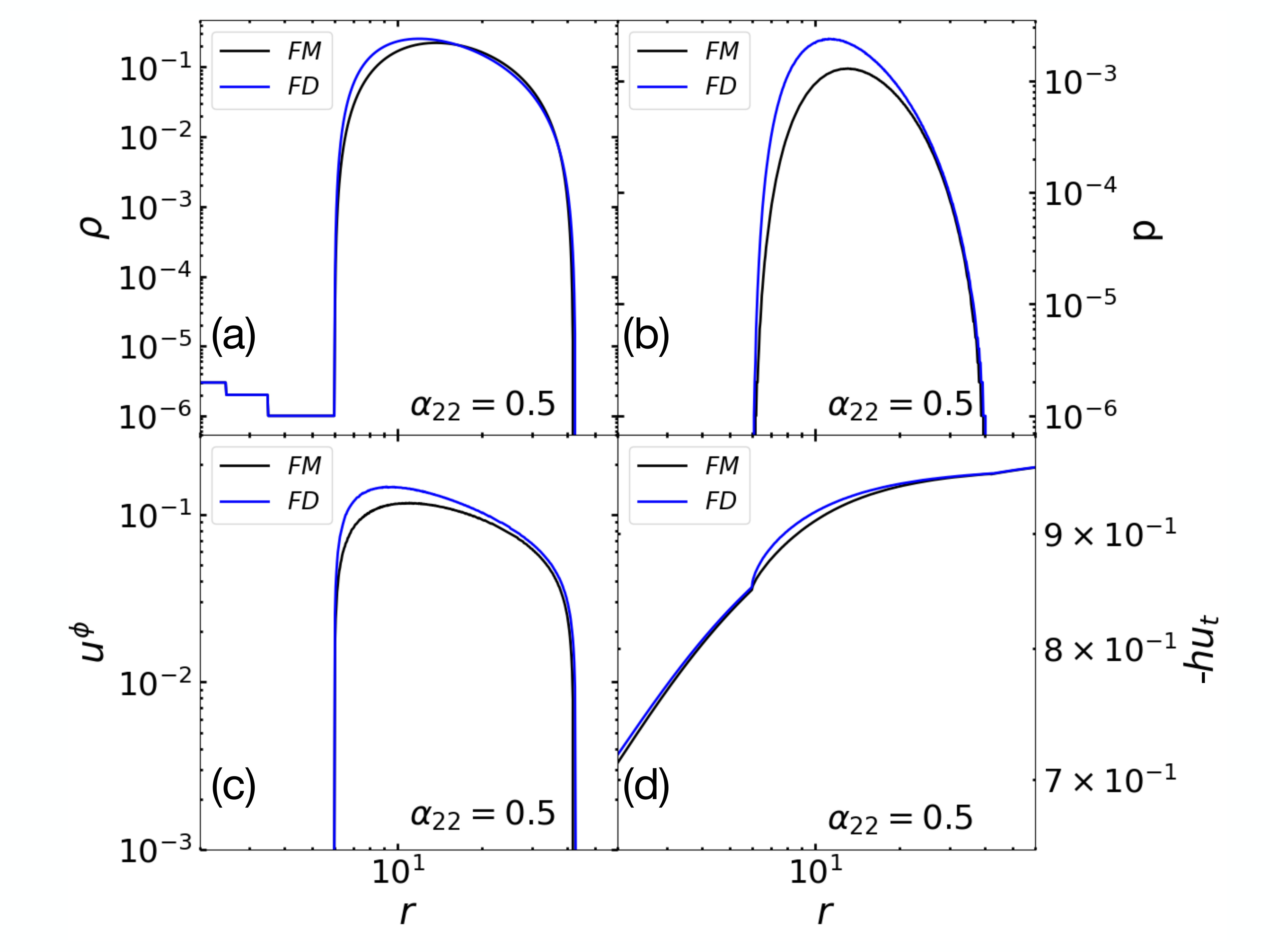}
    \caption{Radial distribution of the vertically integrated density ($\rho$, (a)), pressure ($p$, (b)), azimuthal velocity ($u^\phi$, (c)), and Bernoulli parameter ($-hu_t$, (d)) at $t=0$ for FM (black) and FD torus (blue) in parameterized JP metric with $\alpha_{22}=0.5$.}
    \label{fig: FMFD}
\end{figure}

In Fig.~\ref{fig: FMFD}, we show vertically integrated density ($\rho$), pressure ($p$), azimuthal velocity ($u^\phi$), and Bernoulli parameter ($-hu_t$), respectively. From the comparison, it is seen that the initial torus setup does not vary much. However, some small deviations occur in the interior of the torus. This is due to the different definitions of the constant angular momentum values in both torus solutions. FM torus uses the definition of $l=u_\phi u^t$ (check Appendix A of \cite{Dihingia-etal2024} for $l_*=-u_\phi/u_t$ plots in the case of FM torus), while FD torus uses the definition of $l_*=-u_\phi/u_t$. FD torus has a higher pressure and azimuthal velocity in the middle region of the torus than those of the FM torus. This may influence the structural evolution of the accretion flows in GRMHD simulations. We plan to do such a study in the future and report elsewhere. 

\bibliography{sample631}{}
\bibliographystyle{aasjournal}

\end{document}